\documentclass[preprint,tightenlines]{revtex4}
\def\bwt{\begin{widetext}}
\def\ewt{\end{widetext}}
\def\be{\begin{equation}}
\def\ee{\end{equation}}
\def\bea{\begin{eqnarray}}
\def\eea{\end{eqnarray}}
\def\bean{\begin{eqnarray*}}
\def\eean{\end{eqnarray*}}
\def\bary{\begin{array}}
\def\eary{\end{array}}

\def\bit{\begin{itemize}}
\def\eit{\end{itemize}}

\def\ol{\overline}

\def \pp{^{\prime \prime}}
\begin{document}

\title{\Large{\bf Exploring Final State Hadron Structure \\and
 SU(3) Flavor Symmetry Breaking Effects \\in $D\to PP$ and $D\to PV$ Decays} }
\author{Yue-Liang  Wu}
\author{Ming Zhong} 
\affiliation{Institute of Theoretical Physics, Chinese Academy of Science,\\
P.O.Box 2735, Beijing 100080, P.R. China}
\author{Yu-Feng Zhou}
\affiliation{Institute for Physics, Dortmund University, D-44221,
Dortmund, Germany}

\begin{abstract}
  The nonleptonic two body decays $D\to PP$ and $D\to PV$ are
investigated based on the diagrammatic decomposition in a
generalized factorization formalism. It is shown that to fit the
experimental data, the SU(3) flavor symmetry breaking effects of
the coefficients $a_i$s should be considered in $D\to PP$ decay
modes. In $D\to PV$ decays, the final state hadron structure due
to the pseudoscalar and vector mesons has more important effects
on the coefficients $a_i$s than the SU(3) symmetry breaking
effects. It is found that the nonfactorizable contributions as
well as that of the exchange and annihilation diagrams are
significant in these decays.
\end{abstract}
\preprint{DO-TH 04/04}
\maketitle

\section{Introduction}\label{int}
\vspace{0.3cm}
  Charmed meson nonleptonic two body decays have been
an interesting subject of research \cite{cc,bsw,history,bfbb} for
a long time as it can provide useful information on flavor mixing,
$CP$ violation \cite{cpv} and strong interactions. The theoretical
settlement of this transition type generally appeals to the
factorization hypothesis. Empirically, nonfactorizable corrections
which result from spectator interactions, final state interactions
and resonance effects should be considered. The nonfactorizable
corrections are believed to be significant \cite{fsi}, and they
are relatively hard to be calculated because the charmed quark is
not heavy enough to apply the QCD factorization approach
\cite{bbns} or PQCD approach \cite{lixn} in a reliable manner.
Fortunately, a great number of precise experimental data on
charmed meson nonleptonic two body decays have been accumulated in
recent years. Many new results are expected soon from the
dedicated experiments conducted at BES, CLEO, E791, FOCUS, SELEX
and the two $B$ factories BaBar and Belle. Phenomenological models
based on all kinds of symmetries are of quite importance to guide
the theoretical studies and explore new physics
\cite{kv,rosner,chiang}. But in some cases, the symmetry breaking
effects can be significantly enhanced.

  In the quark diagrammatic scenario, all two-body nonleptonic weak
decays of charmed mesons can be expressed in terms of six distinct
quark-graph contributions \cite{cc,ghlr}: (1) a color-favored tree
amplitude $T$, (2) a color-suppressed tree amplitude $C$, (3) a
W-exchange amplitude $E$, (4) a W-annihilation amplitude $A$, (5)
a horizontal W-loop amplitude $P$ and (6) a vertical W-loop
amplitude $D$. The $P$ and $D$ diagrams play little role in
practice because the CKM matrix elements have the relation
$V^*_{cs}V_{us}\approx -V^*_{cd}V_{ud}$ which will result in
cancellations among these diagrams.

  Based on SU(3) flavor symmetry,
the $T$, $C$, $E$ and $A$ amplitudes were fitted from the measured
D meson decay modes \cite{rosner,chiang}. These amplitudes help
one to understand the generality of charmed meson decays. But
since SU(3) flavor symmetry breaking effects appear to be
important \cite{bre1,bre2}, these fitted data can not describe the
specific properties in certain decay modes. In \cite{zww2003}, we
investigated in detail both the Cabibbo-allowed and singly
Cabibbo-suppressed $D\to PV$ decays based on the diagrammatic
decomposition in the factorization formalism and found that the
SU(3) symmetry breaking effects in the $D\rightarrow PV$ decays
are significant. Two sets of solutions were found in the formalism
of factorization. The case (I) solution can provide satisfactory
explanation in a natural manner on the process $D^+\to \ol
{K}^0K^{*+}$ which is thought to be a puzzle \cite{cl2002}. But
the solution is hard to be explained from the theoretical point of
view because this solution requires such an unexpected large
correction from nonfactorizable contributions that the strong
phase of $a_{T_P}$ has around $150^\circ$ deviation from that of
Wilson coefficients $c_1$. The case (II) solution shows relatively
small correction from nonfactorizable contributions and hence
seems more reasonable from theoretical point of view. But the
solution predicts a relatively small branching ratio of the
process $D^+\to \ol {K}^0K^{*+}$ in comparison with the
experimental result. With such a treatment via solving fifteen
equations for extracting out the same numbers of parameters, it is
hard to consider the experimental uncertainties in \cite{zww2003}.
To investigate what impacts the experimental uncertainties will
bring to the extracted parameters, it is useful to make a
systematic analysis with taking into account the experimental
uncertainties.

  In this paper, we will perform a $\chi^2$ fitting procedure
on charmed mesons decaying to a pseudoscalar and a vector meson
($D\to PV$) and also decaying into two light pseudoscalar mesons
($D\to PP$) by using the quark-graph description based on a
generalized factorization formalism which reflects SU(3) flavor
symmetry breaking effects. Firstly by dividing these diagrams into
factors including SU(3) flavor symmetry breaking effects and
introducing parameters describing the overall properties, we
arrive at two sets of solutions for the parameters from fitting
experimental data. In the view point of diagrammatic
decomposition, the generalized QCD parameters $a_i (i=1,2)$ will
be classified into two sets of parameters $a^P_i$ and $a^V_i$. The
difference between $a^P_i$ and $a^V_i$ arises from the final state
hadron structure of the pseudoscalar and vector mesons in $D\to
PV$. In $D\to PP$ decays, we will show that, to fit the
experimental data, one should classify the parameters $a_i$ into
$a^d_i$ and $a^s_i$, which means that the SU(3) flavor symmetry
breaking effects are important and need to be considered in the
$a_i$s. Thus we can arrive at a conclusion that the coefficients
$a_1$ and $a_2$ depend on either the final state hadron structure
or SU(3) flavor symmetry breaking effects. For $D \to PP$ decay
modes, the SU(3) flavor symmetry breaking effects play an
important role in the coefficients $a_1$ and $a_2$, while for
$D\to PV$ decays, the final state hadron structure becomes more
important for the contributions to the coefficients than the SU(3)
symmetry breaking effect does. The contributions of SU(3) flavor
symmetry breaking effects to $a_1$ and $a_2$ can be neglected in
$D \to PV$ decay modes. Using the fitted parameters as inputs, we
are led to predictions for the branching ratios of other decay
modes which are expected to be measured in the future. In studying
the breaking of the SU(3) symmetry relations, we are able to
quantify the SU(3) breaking effects. The breaking amount of the
SU(3) symmetry relations in some channels can be significant so
that it becomes unreliable to use the SU(3) relations to make
predictions for some decay modes.

  The paper is organized as follows. In section
\ref{sec:notation}, we list the flavor decomposition of the
corresponding mesons and present the quark-diagram description for
the relevant decay modes. In section \ref{sec:formalism}, the
parameterized formalism based on factorization is introduced to
investigate the processes. We then perform a fit procedure in
section \ref{sec:result} to extract the parameters and present
predictions for thirty three $D \to PP$ decay modes and sixty two
$D \to PV$ decay modes. The SU(3) flavor symmetry breaking effects
are discussed in section \ref{sec:breaking}. A short summary and
remark is given in the last section.

\section{Notation And Quark-diagram Formalism}
\label{sec:notation} \vspace{0.3cm}

  We adopt the following quark contents and phase conventions which
have been widely used \cite{rosner,chiang,ghlr,gronau}.
\begin{itemize}
\begin{large}
\item{ {\it Charmed mesons}: $D^0=-c\ol u$, $D^+=c\ol d$,
$D_s^+=c\ol s$;} \item{ {\it Pseudoscalar mesons P}: $\pi^+=u\ol
d$, $\pi^0=(d\ol d-u\ol u)/\sqrt{2}$,
 $\pi^-=-d\ol u$, $K^+=u\ol s$, $K^0=d\ol s$, ${\ol{K}}^0=s\ol d$, $K^-=-s\ol u$,
 $\eta=(-u\ol u-d\ol d+s\ol s)/\sqrt{3}$,
 $\eta^{\prime}=(u\ol u+d\ol d+2s\ol s)/\sqrt{6}$;}
\item{ {\it Vector mesons V}: $\rho^+=u\ol d$, $\rho^0=(d\ol
d-u\ol u)/\sqrt{2}$,
 $\rho^-=-d\ol u$, $\omega=(u\ol u+d\ol d)/\sqrt{2}$, $K^{*+}=u\ol s$,
 $K^{*0}=d\ol s$, ${\ol{K}}^{*0}=s\ol d$, $K^{*-}=-s\ol u$, $\phi=s\ol s$.}
 \end{large}
\end{itemize}
  In the above notations, $u$, $d$ and $s$ quarks transform as a
triplet of flavor SU(3) group, and $-\ol u$, $\ol d$ and $\ol s$
as an antitriplet, so that mesons form isospin multiplets without
extra signs. In general, the $\eta \- \eta^{\prime}$ mixing are
defined as
\begin{equation}\left(\begin{array}{c}\eta \\
\eta^{\prime} \end{array}\right)=\left(\begin{array}{cc} \cos\phi
& -\sin\phi
\\ \sin\phi & \cos\phi \end{array}\right)
\left(\begin{array}{c}\eta_8 \\ \eta_0 \end{array}\right)
\end{equation}
with $\eta_0=(u\ol u+d\ol d+s\ol s)/\sqrt{3}$ and $\eta_8=(-u\ol
u-d\ol d+2s\ol s)/\sqrt{6}$. For convenience, we have taken the
mixing parameter as $\phi=19.5^{\circ}=\sin^{-1}(1/3)$ which is
close to the value $\phi=15.4^{\circ}$ extracted from experiment
\cite{feldmann}.

  The partial width $\Gamma$ for $D\to PP$ and $D\to PV$ decays is expressed in
terms of an invariant amplitude ${\cal A}$. One has
\begin{equation}
\Gamma(D \to PP) = \frac{p}{8 \pi M_D^2}|{\cal A}|^2
\end{equation}
for $D \to PP$ and
\begin{equation}
\Gamma(D \to PV) = \frac{p^3}{8 \pi M_D^2}|{\cal A}|^2
\end{equation}
for $D \to PV$, where
\begin{displaymath}
p=\frac{\sqrt{(M_D^2-(m_1+m_2)^2)(M_D^2-(m_1-m_2)^2)}}{2M_D}
\end{displaymath}
denotes the center-of-mass 3-momentum of each final particle.

  In $D \to PP$ decays, to describe the flavor SU(3) breaking effects,
a subscript $s$ or $d$ is attributed on $T$ and $C$ diagrams to
distinguish the initial $c$ quark transits to $s$ quark or $d$
quark. The subscript $s$ or $d$ is attached to the diagrams $E$
and $A$ dominated by the weak process $c\ol{q}_1 \to q_2\ol{q}_3$
when the antiquark $\ol {q}_3$ is $\ol s$ or $\ol d$. In $D\to PV$
decays, a subscript $P$ or $V$ is assigned to $T$ and $C$, which
are induced by $c\to q_3q_1\ol{q}_2$ with the spectator quark
containing in pseudoscalar or vector final meson. The subscript
$P$ or $V$ is labelled to $E$ and $A$ graphs which are dominated
by the weak process $c\ol{q}_1 \to q_2\ol{q}_3$ when the final
antiquark $\ol {q}_3$ stays in the pseudoscalar or vector meson.
$S$ is added before $E$ or $A$ to distinguish the exchange or
annihilation graph involving in final singlet state contributions
which result from disconnected graphs. The total contributions of
the $SE$ and $SA$ graphs involving in $\pi^0$ and $\rho^0$ mesons
are equal to zero because their contributions resulting from $u\ol
u$ and $-d\ol d$ offset each other due to the isospin SU(2)
symmetry. In the numerical analysis, we will assume that the
contributions of the $SE_P$ and $SE_V$ graphs involving in
$\omega$ and $\phi$ mesons are negligibly small since they seem
not to contradict with the Okubo-Zweig-Iizuka rule. But the
amplitude $SA_V$ seems to play an important role in the $D^+_s\to
\rho^+\eta$ and $D^+_s\to \rho^+\eta^{\prime}$ processes
\cite{chau}. In the ideal mixing case, the process $D^+_s \to
\pi^+\omega$ has the amplitude representation as
$\frac{1}{\sqrt{2}}(A_V+A_P+2SA_P)$. Since $\omega$ has a similar
quark structure in comparison with $\eta$ and $\eta^{\prime}$, we
assume that $SA_P$ has an important contribution in $D^+_s \to
\pi^+\omega$. In the present paper, we shall not consider the
processes which receive contributions from $SA_V$ and $SA_P$
diagrams resulting from the final state particles $\eta$,
$\eta^{\prime}$ or $\omega$. The sign flips in the presentation of
some relevant Cabibbo-favored modes, as well as that of some
doubly Cabibbo-suppressed modes, come from the quark contents of
final light mesons. In the singly Cabibbo-suppressed modes, the
sign flips may come either from the quark contents of the final
light mesons or from the CKM matrix element $V^*_{cd}V_{ud}$ since
$V^*_{cs}V_{us}\approx -V^*_{cd}V_{ud}$ and we choose
$V^*_{cs}V_{us}$ in the calculations. In Table \ref{tab:predict1}
and Table \ref{tab:predict2}, a prime and double prime are added
to the diagrams of singly Cabibbo-suppressed modes and doubly
Cabibbo-suppressed modes respectively to distinguish them from the
Cabibbo-favored ones.

\section{Flavor SU(3) Symmetry Breaking Description In Generalized Factorization Formalism}
\label{sec:formalism} \vspace{0.3cm}

  To investigate the SU(3) flavor symmetry breaking effects, we take
the formalism of a generalized factorization approach
\cite{bsw,gfa}.

  For $D\to PP$ decays, amplitudes can be written in the form as
\begin{eqnarray}
\label{ft}
T_{s,d}&=&\frac{G_F}{\sqrt{2}}V_{q_1q_2}V^*_{cq_3}a_{T_{s,d}}f_{P_1}
(m^2_{D_i}-m^2_{P_2})F^{D_i\rightarrow P_2}_0(m^2_{P_1}), \\
\label{fc}
C_{s,d}&=&\frac{G_F}{\sqrt{2}}V_{q_1q_2}V^*_{cq_3}a_{C_{s,d}}f_{P_1}
(m^2_{D_i}-m^2_{P_2})F^{D_i\rightarrow P_2}_0(m^2_{P_1}), \\
\label{fe}
E_{s,d}&=&\frac{G_F}{\sqrt{2}}V_{q_1q_3}V^*_{cq_2}a_{E_{s,d}}f_{D_i}, \\
\label{fa}
A_{s,d}&=&\frac{G_F}{\sqrt{2}}V_{q_2q_3}V^*_{cq_1}a_{A_{s,d}}f_{D_i}.
\end{eqnarray}

  For $D\to PV$ decays, amplitudes can be written in the form as
\begin{eqnarray}
\label{ftv}
T_V&=&\frac{G_F}{\sqrt{2}}V_{q_1q_2}V^*_{cq_3}a_{T_V}2f_Pm_{D_i}A^{D_i\rightarrow V}_0(m^2_P), \\
\label{ftp}
T_P&=&\frac{G_F}{\sqrt{2}}V_{q_1q_2}V^*_{cq_3}a_{T_P}2f_Vm_{D_i}F^{D_i\rightarrow P}_1(m^2_V), \\
\label{fcv}
C_V&=&\frac{G_F}{\sqrt{2}}V_{q_1q_2}V^*_{cq_3}a_{C_V}2f_Pm_{D_i}A^{D_i\rightarrow V}_0(m^2_P), \\
\label{fcp}
C_P&=&\frac{G_F}{\sqrt{2}}V_{q_1q_2}V^*_{cq_3}a_{C_P}2f_Vm_{D_i}F^{D_i\rightarrow P}_1(m^2_V), \\
\label{fev}
E_{V,P}&=&\frac{G_F}{\sqrt{2}}V_{q_1q_3}V^*_{cq_2}a_{E_{V,P}}2f_{D_i}m_{D_i}, \\
\label{fap}
A_{V,P}&=&\frac{G_F}{\sqrt{2}}V_{q_2q_3}V^*_{cq_1}a_{A_{V,P}}2f_{D_i}m_{D_i}.
\end{eqnarray}
  $D_i$ denotes $D^\pm$, $D_0$ or $D_s$. $F_0$, $F_1$ and
$A_0$ are formfactors defined in the following formalism
\begin{eqnarray}
&&\langle P(p)|\bar q\gamma^{\mu} c|D(p_D) \rangle =\left
[(p_D+p)_{\mu}-\frac{m^2_D-m^2_P}{q^2}q^{\mu}\right
]F_1(q^2)+\frac{m^2_D-m^2_P}{q^2}q^{\mu}F_0(q^2),  \\
&&\langle V(p)|\bar q\gamma^{\mu} (1-\gamma^5) c|D(p_D) \rangle
=-i(m_D+m_V)A_1(q^2)
(\epsilon^{*\mu}-\frac{\epsilon^*\cdot q}{q^2}q^{\mu}) \nonumber \\
&&\hspace{2cm}+i \frac{A_2(q^2)}{m_D+m_V} (\epsilon^{*}\cdot
q)((p_D+p)^{\mu}-\frac{m^2_D-m^2_V}{q^2}q^{\mu})
  -i\frac{2m_V}{q^2} (\epsilon^* \cdot q)A_0(q^2) q^{\mu} \nonumber \\
&&\hspace{2cm} - \frac{2 V(q^2)}{m_D+m_V} \epsilon^{\mu \alpha
\beta \gamma}\epsilon^*_{\alpha} p_{D\beta} p_{\gamma},
\end{eqnarray}
with $q=p_D-p$. $f_P$ and $f_V$ are decay constants defined as
\begin{eqnarray}
&&\langle P(p)|\bar {q}_1\gamma^\mu \gamma_5 q_2|0\rangle
=-if_Pp^{\mu}, \\
&&\langle V(p)|\bar {q}_1\gamma^\mu q_2|0\rangle
=f_Vm_V\epsilon^{\mu}.
\end{eqnarray}

  In naive factorization hypothesis, one has the following
equalities
\begin{eqnarray}
\label{ae1}
&&a_{T_s}=a_{T_d}=a_{T_V}=a_{T_P}=a_1(\mu) , \\
\label{ae2}
&&a_{C_s}=a_{C_d}=a_{C_V}=a_{C_P}=a_2(\mu),
\end{eqnarray}
with
\begin{eqnarray}
\label{a1}
 &&a_1(\mu)=c_1(\mu)+\frac{1}{N_c}c_2(\mu),  \\
 \label{a2}
 &&a_2(\mu)=c_2(\mu)+\frac{1}{N_c}c_1(\mu),
\end{eqnarray}
denoting the relations between quantities $a_{1,2}$ and Wilson
coefficients $c_{1,2}$. $N_c$ is the number of colors. $\mu$ is
the renormalization scale at which $c_1$ and $c_2$ are evaluated.
So $a_1$ and $a_2$ are common real quantities of a certain process
in quark level. To be more explicit, for decay modes induced by
$c\to s$ transition, $a_1$ and $a_2$ are invariant among all modes
in naive factorization hypothesis.

  However, naive factorization approach meets difficulties in
describing all charmed meson decays, particularly for the decay
modes which involve in the color-suppressed diagrams due to the
smallness of $|a_2|$. Furthermore, the coefficients $a_1$ and
$a_2$ in Eqs.(\ref{ae1}) and (\ref{ae2}) depend on the
renormalization scale and $\gamma_5$ scheme at the next to leading
order expansion. It is necessary to consider the nonfactorizable
corrections which involve in hard spectator interactions, final
state interactions and resonance effects etc. In a general case,
one can express $a_1$ and $a_2$ in the form
\begin{eqnarray}
\label{chi1}
&&a_1(\mu)=c_1(\mu)+(\frac{1}{N_c}+\chi_1(\mu))c_2(\mu),  \\
\label{chi2}
&&a_2(\mu)=c_2(\mu)+(\frac{1}{N_c}+\chi_2(\mu))c_1(\mu),
\end{eqnarray}
with $\chi_1(\mu)$ and $\chi_2(\mu)$ terms denoting the
nonfactorizable effects. With nonfactorization corrections the
equalities (\ref{ae1}) and (\ref{ae2}) are not yet satisfied
because each $a_i$ should contain terms from different
corrections. The nonfactorization corrections can also bring phase
differences among these coefficients, and then $a_i$s
($i=T_{s,d,V,P},C_{s,d,V,P}$) become complexes. Currently,
explicit calculations of total nonfactorizable corrections are not
yet possible. In $D \to PV$ decays, we shall take all $a_i$s as
independent complex parameters and assume that the corrections do
not depend on individual decay process at certain scale. In other
words, we do not consider SU(3) flavor symmetry violation
contributions to $a_i$s and it is supposed that mass factors,
decay constants and formfactors have taken on the whole SU(3)
symmetry breaking effects. While in $D \to PP$ decays, the mass
factors, the form factors and decay constants fail to account for
the large SU(3) flavor symmetry breaking effects in $D \to \pi
\pi$, $D \to \pi K$ and $D \to K K$. Nonfactorizable contributions
may cause large SU(3) symmetry breaking \cite{bre2}. Two sets of
coefficients $a^s_i$ and $a^d_i$ are introduced to describe the
SU(3) flavor symmetry breaking effects induced by nonfactorizable
contributions. In both $D\to PP$ and $D\to PV$ decays, the SU(3)
symmetry breaking effects are not considered in the strong phases
in our present analysis.

  The exchange and annihilation diagrams have the following expressions
in naive factorization approach:
\bea \label{fel}
E_{s,d}&=&\frac{G_F}{\sqrt{2}}V_{q_1q_3}V^*_{cq_2}a^{nf}_{E_{s,d}}f_{D_i}
(m^2_{P_1}-m^2_{P_2})F^{P_1P_2}_0(m^2_{D_i}), \\
\label{fal}
A_{s,d}&=&\frac{G_F}{\sqrt{2}}V_{q_2q_3}V^*_{cq_1}a^{nf}_{A_{s,d}}f_{D_i}
(m^2_{P_1}-m^2_{P_2})F^{P_1P_2}_0(m^2_{D_i}),\\
\label{fevl}
E_{V,P}&=&\frac{G_F}{\sqrt{2}}V_{q_1q_3}V^*_{cq_2}a^{nf}_{E_{V,P}}2f_{D_i}m_{D_i}A^{PV}_0(m^2_{D_i}), \\
\label{fapl}
A_{V,P}&=&\frac{G_F}{\sqrt{2}}V_{q_2q_3}V^*_{cq_1}a^{nf}_{A_{V,P}}2f_{D_i}m_{D_i}A^{PV}_0(m^2_{D_i}).
\eea
The formfactors $F^{P_1P_2}_0(m^2_{D_i})$ and
$A^{PV}_0(m^2_{D_i})$ involving in the above formula are not
manifest to relate directly to experimental measurements. The
factorizable contributions of the exchange and annihilation
diagrams are believed to be small. The main contributions of these
diagrams may result from the nonfactorizable forms. Through
intermediate states, these diagrams relate to the tree diagram $T$
and color-suppressed diagram $C$ \cite{cheng2003,zenc}. Their
contributions may be important and can not be ignored. In our
present considerations, we use $a_{E_i, A_i}$($i=s,d,V,P$) defined
in eqs.(\ref{fe}), (\ref{fa}), (\ref{fev}) and (\ref{fap}) as
global parameters to describe mainly the nonfactorizable
contributions. By these definitions, the parameters $a_{E_i,A_i}$
will have two dimensions of energy in $D\to PP$ and will be
dimensionless in $D\to PV$.

\section{Numerical Analysis And Results}
\label{sec:result}

\vspace{0.3cm}

  The explicit evaluation of the relevant formfactors in the
factorization formula (\ref{ft}), (\ref{fc}) and
(\ref{ftv})-(\ref{fcp}) is a hard task because of the
nonperturbative long distance effects of QCD. Various methods,
such as QCD sum rules\cite{kr,ww}, lattice simulations
\cite{flynn,abada} and phenomenological quark model
\cite{isgur,melikhov}, have been developed to estimate the long
distance effects to rather high certainties. The formfactors of
$D$ mesons decaying to light mesons have been widely discussed in
\cite{bbs,ms2000,gus,fs1998,ball,wwz}. In our present
considerations, we shall use the results of form factors obtained
by Bauer, Stech and Wirbel \cite{bsw,bbs} based on the quark
model. They have been found to be rather successful in describing
a number of processes concerning $D$ mesons. The values of the
relevant formfactors evaluated at $q^2=0$ are listed in Table
\ref{tab:formfactors}. For the dependence on $q^2$, the
formfactors are assumed to behave as a monopole dominance
\begin{eqnarray}
D\rightarrow P:&& \quad F_0(q^2)=\frac{F_0(0)}{1-q^2/m^2_{F^{**}}}, \\
&&\quad F_1(q^2)=\frac{F_1(0)}{1-q^2/m^2_{F^*}}, \\
D\rightarrow V:&& \quad A_0(q^2)=\frac{A_0(0)}{1-q^2/m^2_F},
\end{eqnarray}
where $m_F$, $m_{F^*}$ and $m_{F^{**}}$ are the pole masses given
in Table \ref{tab:formfactors}.

  It is noted that the formfactors are more appropriate to be viewed as the relative
scaling factors that characterize one source of SU(3) flavor
symmetry breaking effects in hadronic matrix elements since we
take the $a_i$s as free parameters that need to be extracted from
experimental inputs in the present method. The relative ratio
between the formfactors is what we really care about.

  The input values for the light pseudoscalar and vector decay constants
are presented in Table \ref{tab:decayconstant} \cite{akl,ns}.
These values generally coincide with experiments. The decay
constants $f^u_{\eta}$, $f^s_{\eta}$, $f^u_{\eta^{\prime}}$ and
$f^s_{\eta^{\prime}}$ involving in factorization formula should be
defined as follow \cite{akl}
\begin{eqnarray}
&&\langle 0|\ol u\gamma^\mu \gamma_5 u|\eta^{(\prime)}(p)\rangle
=if^u_{\eta^{(\prime)}}p^{\mu} , \\
&&\langle 0|\ol s\gamma^\mu \gamma_5 s|\eta^{(\prime)}(p)\rangle
=if^s_{\eta^{(\prime)}}p^{\mu}.
\end{eqnarray}
Then the quantities $f^u_{\eta}$, $f^s_{\eta}$,
$f^u_{\eta^{\prime}}$ and $f^s_{\eta^{\prime}}$ take the formalism
\begin{eqnarray}
&&f^u_{\eta}=\frac{f_8}{\sqrt{6}}\cos
\phi+\frac{f_0}{\sqrt{3}}\sin \phi ,\\
&&f^s_{\eta}=\frac{2f_8}{\sqrt{6}}\cos
\phi-\frac{f_0}{\sqrt{3}}\sin \phi ,\\
&&f^u_{\eta^{\prime}}=-\frac{f_8}{\sqrt{6}}\sin
\phi+\frac{f_0}{\sqrt{3}}\cos \phi ,\\
&&f^s_{\eta^{\prime}}=\frac{2f_8}{\sqrt{6}}\sin
\phi+\frac{f_0}{\sqrt{3}}\cos \phi .
\end{eqnarray}
Making use of these definitions, the following factorization
formalisms are adopted in the $D\to \eta(\eta^{\prime})V$
transition calculation
\begin{eqnarray}
2C_V(D_i\to \eta
V)&=&\frac{G_F}{\sqrt{2}}V_{q_1q_2}V^*_{cq_3}a_{C_V}2(f^u_{\eta}+f^s_{\eta})
m_{D_i}A^{D_i\rightarrow V}_0(m^2_{\eta}), \\
C_V(D_i\to \eta^{\prime}
V)&=&\frac{G_F}{\sqrt{2}}V_{q_1q_2}V^*_{cq_3}a_{C_V}2(f^s_{\eta^{\prime}}-f^u_{\eta^{\prime}})
m_{D_i}A^{D_i\rightarrow V}_0(m^2_{\eta^{\prime}}).
\end{eqnarray}

  The other parameters used in the numerical calculation are the
masses of relevant mesons, lifetimes of charmed mesons and
relevant CKM matrix elements. We adopt the relevant values given
in \cite{PDG2002}.

  For convenience, we may express the complex parameters $a_i$ as
\begin{equation}
a_i=|a_i|e^{i\delta_{a_i}}.
\end{equation}
The $\delta_{a_i}$s characterize the strong phases. One can always
choose $\delta_{a_{T_s}}=0$ in $D\to PP$ and $\delta_{a_{T_V}}=0$
in $D\to PV$ so that all the other strong phases are relative to
$\delta_{a_{T_s}}$ and $\delta_{a_{T_V}}$. There are $15$
independent parameters to be extracted from experiments in both
$D\to PP$ and $D\to PV$.

  To conduct a fit procedure, we construct a $\chi^2$ function
which has the following form
\be \chi^2=\sum\limits_j \frac{(f_j(a_i)-\langle
f_j\rangle)^2}{\sigma^2_j} \ee
where $\langle f_j\rangle$ and $\sigma_j$ are the central values
and corresponding errors of the experimentally measured
observables. $f_j(a_i)$ are the theoretical expressions for the
observables. They are the functions of parameters $a_i$s. The set
of $a_i$s which minimizes the $\chi^2$ function will be regarded
as the best estimated values.

  There are $17$
experimental data points for $13$ parameters ($a^s_A$ does not
appear in these experimental data) and $22$ data points for $15$
parameters, as shown in Table \ref{tab:predict1} and Table
\ref{tab:predict2} respectively. We list in Table \ref{tab:para}
the parameters with $1\sigma$ errors obtained in our present
analysis. FIT $\alpha$ and FIT A are obtained without any
constraint to the parameters. A large $|a^s_2/a^d_2|\approx 2.0$
ratio predicted by FIT $\alpha$ is an indiction of inscrutably
large flavor SU(3) breaking effects. Constraining the ratio to the
smallest extent, we get FIT $\beta$ with the ratio
$|a^s_2/a^d_2|\approx 1.1$. By "the smallest extent", we mean
that, if we continue to suppress the ratio down, the predicted
branching ratios of some decay modes in Table \ref{tab:predict1}
will be inconsistent with the experimental data. FIT A predicts an
unusually large ratio $|a^V_2/a^V_1| \approx 1.1$ which indicates
that the nonfactorizable contributions to $a^V_2$ are of great
importance. By constraining the value of $|a^V_2|$ to be as small
as possible, we obtain FIT B with the ratio $0.9$. The next
leading order Wilson coefficients $c_1(m_c)=1.174$ and
$c_2(m_c)=-0.356$ in the naive dimensional regularization (NDR)
scheme or $c_1(m_c)=1.216$ and $c_2(m_c)=-0.424$ in the
'tHooft-Veltman (HV) scheme are given in Ref.\cite{bbl} when
$\Lambda_{\ol {MS}}=0.215$GeV. The present relatively large values
of $|a_2^s|$, $|a_2^d|$, $|a_2^V|$ and $|a_2^P|$ can not be
explained from formula (\ref{a2}). They imply that nonfactorizable
contributions are of significance in both $D\to PP$ and $D\to PV$
decays. To fit the experimental result of $Br(D^0\to
K^0\ol{K}^0)=(0.071\pm0.019)\%$, $a_E^s$ should differ much from
$a_E^d$. In $D\to PV$ decays, because we have considered the
errors of experimental data in the $\chi^2$ fit and used more
experimental results as constraints, the present resulting
parameters appear more reasonable than that of case (I) solution
presented in \cite{zww2003}, as the strong phases of the
parameters $a_1^{P,V}$ and $a_2^{P,V}$ are not in contradiction to
that predicted from QCD.

  We present the resultant predictions for a variety of charmed meson
decay processes in Table \ref{tab:predict1} for $D\to PP$ decays
and in Table \ref{tab:predict2} for $D\to PV$ decays. Note that
there are no enough experimental data to extract the parameter
$a^s_A$ in $D\to PP$ decays. To make predictions for some relevant
decay modes which receive contribution from $A_s$ diagram, we take
the assumption $a^s_A=a^d_A$. For a comparison, we also list the
results obtained in Ref.\cite{lp2002}. The predictions for a
number of singly and doubly Cabibbo-suppressed modes can be used
to test our present analysis in the near future.

  Note that in the assumption of $SA_P=0$, we have the branching ratio
$14\%$ for the process $D^+_s \to \pi^+\omega$, which is much
larger than the experimental result $(0.28\pm 0.11)\%$. To
accommodate the experimental data, significant contribution from
$SA_P$ diagram, i.e. $SA_P \sim -A_p$, should be introduced
\cite{zww2003}.

\section{SU(3) Flavor Symmetry Breaking}
\label{sec:breaking}

\vspace{0.3cm}

   As pointed out in Ref.\cite{bre1,bre2}, SU(3) breaking effects in
charmed meson decays appear to be important. The violation may
come from the finite strange quark mass, the final state
interactions and resonances. In the SU(3) flavor symmetry limit,
there are a number of relations among different decay modes. Based
on the above extracted values for the parameters, we can discuss
how large are the SU(3) breaking effects in $D\rightarrow PP$ and
$D\rightarrow PV$ decays.

   We present these relations in Table \ref{tab:breaking1} for
$D\rightarrow PP$ and Table \ref{tab:breaking2} for $D\rightarrow
PV$. The left hand side(LHS) values of the relations whose
deviation from unit represents the breaking amounts of SU(3)
flavor symmetry relations are listed in the second columns.

  It is noted that though these relations
deviating from unit reflects the SU(3) flavor symmetry breaking
effects, the ones composed of three decay modes and those composed
of two decay modes have different sources of breaking terms. To be
clear, we take the expressions $\frac{|\lambda{\cal A}(D^+\to
\pi^+\ol{K}^{*0})+\sqrt{2}{\cal A}(D^+\to
\pi^+\rho^0)|}{|\lambda\sqrt{2}{\cal A}(D^+_s\to \pi^+\rho^0)|}$
and $\frac{|{\cal A}(D^0\to K^+K^{*-})|}{|{\cal A}(D^0\to
\pi^+\rho^-)|}$ as examples.
\begin{eqnarray}
\label{b1}
&&\frac{|\lambda{\cal A}(D^+\to
\pi^+\ol{K}^{*0})+\sqrt{2}{\cal A}(D^+\to
\pi^+\rho^0)|}{|\lambda\sqrt{2}{\cal A}(D^+_s\to
\pi^+\rho^0)|}\nonumber \\
=&&\frac{|(T_V+C_P)(D^+\to
\pi^+\ol{K}^{*0})-(T_V+C_P-A_P+A_V)(D^+\to
\pi^+\rho^{0})|}{|A_V(D^+_s\to \pi^+\rho^{0})-A_P(D^+_s\to
\pi^+\rho^{0})|},\\
\label{b2}
&&\frac{|{\cal A}(D^0\to K^+K^{*-})|}{|{\cal A}(D^0\to
\pi^+\rho^-)|}=\frac{|T_V(D^0\to K^+K^{*-})+E_P(D^0\to
K^+K^{*-})|}{|-T_V(D^0\to \pi^+\rho^-)-E_P(D^0\to \pi^+\rho^-)|}.
\end{eqnarray}
In the limit of SU(3) flavor symmetry, we have the following
relations
\begin{eqnarray}
\label{c1}
&&T_V(D^+\to \pi^+\ol{K}^{*0})=T_V(D^+\to \pi^+\rho^{0}),\\
\label{c2}
&&C_P(D^+\to \pi^+\ol{K}^{*0})=C_P(D^+\to \pi^+\rho^{0}),\\
\label{c3}
&&A_V(D^+\to \pi^+\rho^{0})=A_V(D^+_s\to \pi^+\rho^{0}),\\
\label{c4}
&&A_P(D^+\to \pi^+\rho^{0})=A_P(D^+_s\to \pi^+\rho^{0}),\\
\label{c5}
&&T_V(D^0\to K^+K^{*-})=T_V(D^0\to \pi^+\rho^-) ,   \\
\label{c6}
&&E_P(D^0\to K^+K^{*-})=E_P(D^0\to \pi^+\rho^-),
\end{eqnarray}
which make the ratios eq.(\ref{b1}) and eq.(\ref{b2}) equal to
one. But from formula (\ref{ftv})-(\ref{fap}), one can find that
relations in eqs. (\ref{c1})-(\ref{c6}) are in general not valid.
The different masses of the charmed mesons and the final light
mesons, and the different values of formfactors and decay
constants can break the relations in eqs. (\ref{c1})-(\ref{c6}),
and thus break the SU(3) flavor symmetry relations in (\ref{b1})
and (\ref{b2}). In addition, by comparing with (\ref{b1}) and
(\ref{b2}), one can see that the relations concerning only two
decay modes represent the relative SU(3) flavor symmetry breaking
amounts of the same diagrams which we call the main diagrams for
convenience in later use, while the relations consisting of three
decay modes contain additional SU(3) flavor symmetry breaking
effects from the other diagrams. So in the relations containing
three decay modes, if the SU(3) flavor symmetry breaking
contributions of the other diagrams have comparable amounts in
comparison with the main diagrams, then the relations will be
broken down badly. The main diagrams $|T+C|$ in $D^+\to
\pi^+\ol{K}^0$, $|A_V-A_P|$ in $D^+_s\to \pi^+\rho^0$ and
$|T_V+C_P|$ in $D^+\to \pi^+\ol{K}^{*0}$ are relatively small,
which usually leads to a significant breaking for the relations
when taking $A(D^+\to \pi^+\ol{K}^0)$, $A(D^+_s\to \pi^+\rho^0)$
and $A(D^+\to \pi^+\ol{K}^{*0})$ as denominators. We present
explicitly some of these relations calculated from the parameters
of FIT $\alpha$ and FIT A as follows:
\begin{eqnarray}
&&\frac{|\sqrt{2}\kappa{\cal A}(D^+\to K^+\pi^0)-\kappa{\cal
A}(D^+\to K^0\pi^+)|}{|\lambda{\cal A}(D^+\to
\ol{K}^0\pi^+)|}=2.21 ,\\
&&\frac{|\lambda{\cal A}(D^+\to \pi^+\ol{K}^{*0})+\sqrt{2}{\cal
A}(D^+\to \pi^+\rho^0)|}{|\lambda\sqrt{2}{\cal A}(D^+_s\to
\pi^+\rho^0)|}=0.60 ,\\
&&\frac{|\lambda{\cal A}(D^+_s\to \ol {K}^0K^{*+})+\sqrt{2}{\cal
A}(D^+_s\to K^{*+}\pi^0)|}{|\lambda\sqrt{2}{\cal A}(D^+_s\to
\pi^+\rho^0)|}=1.42,\\
&&\frac{|\lambda{\cal A}(D^+_s\to K^+\ol {K}^{*0})+{\cal A}(D^+\to
K^+\ol {K}^{*0})|}{|\lambda{\cal A}(D^+\to \pi^+\ol
{K}^{*0})|}=0.58 .
\end{eqnarray}
It is obvious that SU(3) flavor symmetry analysis is not
applicable to such processes.

  Besides the mass factors, the
form factors and decay constants, one should also consider the
contributions of $a_i$ factors when studying the SU(3) symmetry
breaking effects in $D\to PP$ decay modes. The situations are more
complicated than that in $D\to PV$ decay modes. General speaking,
the SU(3) flavor symmetry breaking effects are more important in
$D\to PP$ decays. The first two relations in Table
\ref{tab:breaking1} and Table \ref{tab:breaking2} still conserve
because all the decay modes in them form an isospin triangle
respectively.

\section{Summary And Conclusion}
\label{sec:summary}
\vspace{0.3cm}
  We have performed a $\chi^2$ fitting analysis on the
$D\to PP$ and $D\to PV$ decays in the formalism of the
factorization hypotheses. To fit the experimental data, it is
vital to consider the SU(3) flavor symmetry breaking effects of
the coefficients $a_i$s in $D\to PP$ decay modes. In $D\to PV$
decays, the final state hadron structure of the pseudoscalar and
vector mesons has more important impact on the coefficients $a_i$s
than the SU(3) symmetry breaking effects. The nonfactorizable
contributions, as well as that of the exchange and annihilation
diagrams, are found to be important in these decays. In the
formalism of the relations obtained in the SU(3) symmetry limit,
the total SU(3) symmetry breaking amount of certain processes in
$D\to PP$ can reach $120\%$ when the three symmetry breaking
effects due to $a_i$ factors, mass factors and due to form factors
and decay constants become to be coherently added. The total
breaking amount of some processes in $D\to PV$ can add up to
$50\%$. The breaking amount of the SU(3) symmetry relations in
some channels is so significant that it becomes unreliable to use
the SU(3) relations to make predictions for some decay modes. More
precise measurement on the process $D^+\to \ol {K}^0K^{*+}$ is
important for understanding the SU(3) symmetry breaking effects
and nonfactorizable contributions. As an independent check, it is
useful to measure the process $D^+_s\to K^0 \rho^+$. The processes
$D^0\to \pi^+\rho^-$, $D^0\to \pi^-\rho^+$, $D^0\to \pi^0\rho^0$,
$D^+\to \pi^+\omega$, $D^+\to \pi^0 \rho^+$, $D^+\to K^0\rho^+$,
$D^+\to \pi^0K^{*+}$, $D^+_s\to K^+\omega$ and $D^+_s\to
\pi^0K^{*+}$ are predicted to be at the experimental sensitivity.
It is expected to explore the final hadron structure and SU(3)
flavor symmetry breaking effects in $D\to PP$ and $D\to PV$ decays
in BES, CLEO-c, BaBar and Belle.

\section*{ACKNOWLEDGMENTS}

  This work was supported in part by the key projects of Chinese
Academy of Sciences, the National Science Foundation of China
(NSFC), the BEPC National Lab Opening Project and Associate
Scheme at Abdus Salam ICTP, Italy.

\newpage
\begin{table}[hbp]
\caption{\small Relevant formfactors at zero momentum transfer for
$D\to P$ and $D\to V$ transitions and pole masses in BSW model.
\label{tab:formfactors}} \vspace{8pt}
\begin{scriptsize}
\begin{center}
\begin{tabular}{| c | c | c | c | c | c | c | c | c | c | }
\hline Decay & $D\to \pi$ & $D\to \rho(\omega)$ & $D\to K$ & $D\to
K^*$ & $D_s\to K$ & $D_s\to K^*$ & $D_s\to \phi$ &$D\to \eta/\eta^{\prime}$ &$D_s\to \eta/\eta^{\prime}$\\
\hline
 $F_1$& 0.692 &  & 0.762 &  & 0.643 &  &   & 0.681/0.655 & 0.723/0.704    \\
\hline
 $A_0$ &       & 0.669 &       & 0.733 &       & 0.634 & 0.700  &  &     \\
\hline
 $m_F(GeV)$ &       & 1.87  &      & 1.97  &    & 1.87 & 1.97   &             &        \\
\hline
$m_{F^*}(GeV)$& 2.01 &      & 2.11 &       & 2.01 &    &        &   2.01      &   2.11  \\
\hline
$m_{F^{**}}(GeV)$& 2.47 &      & 2.60 &       & 2.47 &    &        &   2.47      &   2.60\\
\hline
\end{tabular}
\end{center}
\end{scriptsize}
\end{table}
\begin{table}[hbp]
\caption{\small Values of decay constants in MeV.
\label{tab:decayconstant}} \vspace{8pt}
\begin{center}
\begin{tabular}{| c | c | c | c | c | c | c | c | c | c | c | c |}
\hline $f_{\pi}$& $f_{K}$ & $f_{8}$ & $f_{0}$ & $f_D$ & $f_{D_s}$&
$f_{\rho}$
& $f_{K^*}$ & $f_{\omega}$ & $f_{\phi}$ & $f_{D^*}$ & $f_{D^*_s}$\\
\hline
 134 & 158 &  168 & 157 & 200 & 234 & 210 & 214 & 195 & 233 & 230 & 275 \\
\hline
\end{tabular}
\end{center}
\end{table}
\begin{table}[hbp]
\caption{\small Parameters $a_i$s fitted from experimental data at
$1 \sigma$ errors. The first entry is for amplitude and the second
entry for the strong phase. $a^{s,d,V,P}_1$ and $a^{s,d,V,P}_2$
denote $a_{T_{s,d,V,P}}$ and $a_{C_{s,d,V,P}}$
respectively.\label{tab:para}} \vspace{8pt}
\begin{center}
\begin{tabular}{|c|c|c|c|c|c|}
\hline  \multicolumn{3}{|c|}{\hspace*{1.5cm} $D \rightarrow
PP$ \hspace*{1.5cm}} & \multicolumn{3}{|c|}{\hspace*{1.5cm} $D \rightarrow PV$ \hspace*{1.5cm}}\\
\hline   & FIT $\alpha$  & FIT $\beta$&  &FIT A& FIT B \\
$\chi^2/d.o.f.$& 4.06/4   & 8.16/4  & & 8.22/7  &10.30/7   \\
\hline \hline
$a^s_1$&$1.08\pm 0.04$&$1.10\pm 0.03$&$a^V_1$&$1.13\pm 0.08$&$1.10\pm 0.07$\\
        &$---$&$---$& &$---$&$---$\\
        \hline
 $a^d_1$&$1.04\pm 0.09$&$1.09\pm 0.09$&$a^P_1$&$1.29\pm 0.04$&$1.29\pm 0.04$\\
        &$(8.73\pm 7.96)^\circ$&$(11.98\pm 7.85)^\circ$& &$(10.04\pm 16.62)^\circ$&$(-1.36\pm 13.52)^\circ$\\
        \hline \hline
$a^s_2$&$-0.73\pm 0.04$&$-0.73\pm 0.04$&$a^V_2$&$-1.19\pm 0.06$&$-1.00\pm 0.05$\\
        &$(-26.76\pm 1.60)^\circ$&$(-26.25\pm 1.55)^\circ$& &$(-11.09\pm 20.01)^\circ$&$(-10.74\pm 10.31)^\circ$\\
        \hline
 $a^d_2$&$-0.36\pm 0.20$&$-0.65\pm 0.06$&$a^P_2$&$-0.78\pm 0.03$&$-0.77\pm 0.02$\\
        &$(-53.40\pm 28.65)^\circ$&$(-35.12\pm 14.50)^\circ$& &$(-21.75\pm 1.38)^\circ$&$(-22.15\pm 2.23)^\circ$\\
  \hline \hline
$a^s_E(GeV^2)$&$0.24\pm 0.11$&$0.25\pm 0.11$&$a^V_E$&$0.07\pm 0.03$&$0.26\pm 0.04$\\
        &$(-49.43\pm 20.63)^\circ$&$(-58.75\pm 19.48)^\circ$& &$(-166.87\pm 50.96)^\circ$&$(-115.50\pm 12.78)^\circ$\\
        \hline
 $a^d_E(GeV^2)$&$1.01\pm 0.08$&$1.01\pm 0.08$&$a^P_E$&$0.51\pm 0.02$&$0.50\pm 0.03$\\
        &$(-120.94\pm 2.92)^\circ$&$(-122.02\pm 2.82)^\circ$& &$(82.82\pm 4.01)^\circ$&$(78.55\pm 5.73)^\circ$\\
   \hline \hline
$a^s_A(GeV^2)$&$---$&$---$&$a^V_A$&$0.52\pm 0.05$&$0.53\pm 0.05$\\
        &$---$&$---$& &$(-76.66\pm 25.78)^\circ$&$(-78.24\pm 26.13)^\circ$\\
        \hline
 $a^d_A(GeV^2)$&$0.43\pm 0.09$&$0.46\pm 0.09$&$a^P_A$&$0.59\pm 0.03$&$0.60\pm 0.02$\\
        &$(90.04\pm 13.75)^\circ$&$(89.94\pm 14.27)^\circ$& &$(-76.29\pm 41.25)^\circ$&$(-77.93\pm 31.52)^\circ$\\
  \hline
\end{tabular}
\end{center}
\end{table}
\begin{table}[hbp]
\caption{\small Predicted branching ratios for charmed mesons
decaying to two pseudoscalar mesons. Single prime and double
primes are added to the representations to denote the singly
Cabibbo-suppressed processes and doubly Cabibbo-suppressed
processes. $C_{s_1}$ and $C_{s_2}$, as well as $C_{d_1}$ and
$C_{d_2}$, result from the exchange of the final mesons.
\label{tab:predict1}} \vspace{8pt}
\begin{center}
\begin{tabular}{|c|c|c|c|c|c|c|}
\hline  \multicolumn{2}{|c|}{Decay Modes}& Representation
     & Experimental  & \multicolumn{2}{|c|}{Present ${\cal B}\times 10^{-2}$}&LP\cite{lp2002}  \\
     \cline{5-6}
\multicolumn{2}{|c|}{}& &${\cal B}\times 10^{-2}$ &FIT
$\alpha$&FIT $\beta$
    &${\cal B}\times 10^{-2}$\\
    \hline
  & $K^- \pi^+$ & $T_s + E_d$
     & $3.80 \pm 0.09$ &3.79& 3.80 &3.847\\
 & $\ol{K}^0 \pi^0$ & $\frac{1}{\sqrt{2}}(C_s-E_d)$
     & $2.28 \pm 0.22$ &2.27& 2.24 & 1.310\\
  &$\ol{K}^0 \eta$ & $\frac{1}{\sqrt{3}}C_s$
     & $0.76 \pm 0.11$ &0.80& 0.81 & \\
 & $\ol{K}^0 \eta\,'$ & $-\frac{1}{\sqrt{6}}(C_s+3E_d)$
     & $1.87 \pm 0.28$ & 1.85&1.88&  \\
  &$\pi^+ \pi^-$ & $-(T\,'_d+E\,'_d)$
     &$0.143 \pm 0.007$&0.144& 0.144 & 0.151\\
 & $\pi^0 \pi^0$ &  $-\frac{1}{\sqrt{2}}(C\,'_d-E\,'_d)$
     &$0.084 \pm 0.022$ &0.078& 0.097 & 0.115\\
 & $K^+ K^-$ &  $T\,'_s+E\,'_s$
     & $0.412 \pm 0.014$&0.413&0.413 & 0.424 \\
 & $K^0 \ol{K}^0$ & $E\,'_s-E\,'_d$
    & $0.071 \pm 0.019$ &0.069&0.062 & 0.130\\
$D^0$&$K^+\pi^-$& $-(T\pp_d+E\pp_s)$&$0.0148 \pm 0.0021$&0.0150&0.0151 & 0.033  \\
 & $\eta \pi^0$ &{\small
  $\frac{1}{\sqrt{6}}(C\,'_s+C\,'_{d_1}-C\,'_{d_2}-2E\,'_d-SE\,')$}
     & --- &0.069&0.068  & \\
  &$\eta\,' \pi^0$ &{\scriptsize
  $\frac{1}{\sqrt{12}}(2C\,'_s-C\,'_{d_1}+C\,'_{d_2}+2E\,'_d+4SE\,')$}
     & --- &0.088& 0.091& \\
  &$\eta \eta$ &{\small
$\frac{1}{3\sqrt{2}}(2C\,'_s+2C\,'_d-2E\,'_s+2E\,'_d+4SE)$}
     & ---&0.011&0.016& \\
 & $\eta \eta\,'$ &{\tiny
  $\frac{1}{\sqrt{18}}(2C\,'_{s_1}-C\,'_{s_2}-C\,'_{d_1}-C\,'_{d_2}-4E\,'_s-2E\,'_d-7SE)$}
     & ---&0.026&0.030 &  \\
   & $K^0\pi^0$&$ -\frac{1}{\sqrt{2}}(C\pp_d-E\pp_s)$&---&0.002&0.005 & 0.008\\
  &$K^0 \eta$ & $-\frac{1}{\sqrt{3}}(C\pp_d-E\pp_s+SE\pp)$
     &---&0.001&0.002 & \\
 & $K^0 \eta\,'$ & $\frac{1}{\sqrt{6}}(C\pp_d+3E\pp_s+4SE\pp)$
     &---&0.0&0.0&\\
\hline  &$\ol{K}^0 \pi^+$ & $T_s+C_s$
     & $2.77 \pm 0.18$&2.76&2.76& 2.939 \\
  &$\pi^+ \pi^0$  &$-\frac{1}{\sqrt{2}}(T\,'_d+C\,'_d)$& $0.25 \pm
  0.07$&0.25
     &0.19&0.185\\
   &$\eta\pi^+$&{\scriptsize $\frac{1}{\sqrt{3}}(T\,'_d+C\,'_s+C\,'_d+2A\,'_d+SA\,')$}&$0.30 \pm 0.06$&0.34&0.37&\\
  &$\eta'\pi^+$&{\scriptsize$-\frac{1}{\sqrt{6}}(T\,'_d-2C\,'_s+C\,'_d+2A\,'_d+4SA\,')$}&$0.50 \pm 0.10$&0.45&0.42&
  \\
 $D^+$ &$K^+ \ol{K}^0$ & $T\,'_s-A\,'_d$
     & $0.58 \pm 0.06$&0.62 &0.62 & 0.764\\
  &$K^0\pi^+$&$ -(C\pp_d+A\pp_s)$&--- &0.012&0.026 & 0.053\\
   &$K^+\pi^0$&$-\frac{1}{\sqrt{2}}(T\pp_d-A\pp_s)$&---&0.021&0.023&0.055 \\
   &$K^+\eta$&$\frac{1}{\sqrt{3}}(T\pp_d+SA\pp)$&---&0.011&0.012
  & \\
  &$K^+\eta'$&$-\frac{1}{\sqrt{6}}(T\pp_d+3A\pp_s+4SA\pp)$&---
  &0.005&0.006& \\
\hline
 & $\ol{K}^0 K^+$ & $C_s+A_d$
     & $3.6 \pm 1.1$&3.06&3.13&4.623\\
  &$\pi^+\eta$&$\frac{1}{\sqrt{3}}(T_s-2A_d-SA)$&$1.7 \pm 0.5$&1.05&1.09&1.131 \\
  &$\pi^+\eta\,'$&$\frac{2}{\sqrt{6}}(T_s+A_d+2SA)$&$3.9 \pm 1.0$&4.19&4.43& \\
$D^+_s$  &$\pi^+ K^0$ &  $-(T\,'_d-A\,'_s)$
     & $<0.8$&0.24&0.26&0.373 \\
 & $\pi^0 K^+$ & $-\frac{1}{\sqrt{2}}(C\,'_d+A\,'_s)$
     & ---&0.047&0.090&0.146\\
 & $\eta K^+$&$\frac{1}{\sqrt{3}}(T\,'_s+C\,'_s+C\,'_d-SA\,')$&---&0.055&0.040&0.300\\
 & $\eta\,' K^+$&{\scriptsize $\frac{1}{\sqrt{6}}(2T\,'_s+2C\,'_s-C\,'_d+3A\,'_s+4SA\,')$}&
  ---&0.090&0.102&\\
 & $K^+K^0$&$-(T\pp_d+C\pp_d)$&---&0.014&0.010 & 0.012\\
\hline
\end{tabular}
\end{center}
\end{table}
\begin{table}[hbp]
\caption{\small Predicted branching ratios for charmed mesons
decaying to one pseudoscalar and one vector meson. Single prime
and double primes are added to the representations to denote the
singly Cabibbo-suppressed processes and doubly Cabibbo-suppressed
processes. \label{tab:predict2}} \vspace{8pt}
\begin{center}
\begin{tabular}{|c|c|c|c|c|c|c|}
\hline \multicolumn{2}{|c|}{Decay Modes}& Representation
     & Experimental &\multicolumn{2}{|c|}{Present ${\cal B} (\times 10^{-2})$}& LP\cite{lp2002}\\
     \cline{5-6}
 \multicolumn{2}{|c|}{}& &${\cal B} (\times 10^{-2})$&FIT A&FIT B&${\cal B} (\times 10^{-2})$\\
  \hline
  & $K^{*-} \pi^+$ & $T_V + E_P$
     & $6.0 \pm 0.5$&5.93 &5.97  & 4.656\\
  &$K^- \rho^+$ & $T_P + E_V$
     & $10.2 \pm 0.8$&9.99 &9.90 & 11.201\\
  &$\ol{K}^{*0} \pi^0$ & $\frac{1}{\sqrt{2}}(C_P - E_P)$
     & $2.8 \pm 0.4$&2.72&2.81&3.208\\
  &$\ol{K}^0 \rho^0$ & $\frac{1}{\sqrt{2}}(C_V - E_V)$
     & $1.47 \pm 0.29$ &1.49&1.25& 0.759 \\
  &$\ol{K}^{*0} \eta$ & $\frac{1}{\sqrt{3}}(C_P + E_P - E_V+SE_V)$
     & $1.8 \pm 0.4$ &1.50&1.94 & \\
  &$\ol{K}^0 \omega$ & $-\frac{1}{\sqrt{2}}(C_V + E_V)$
     & $2.2 \pm 0.4$&2.11&1.80&1.855\\
  &$\ol{K}^0 \phi$ & $-E_P-SE_P$
     & $0.94 \pm 0.11$&0.95 &0.90 & \\
  &$K^+ K^{*-}$ &  $T\,'_V+E\,'_P$
     & $0.20\pm0.11$&0.25&0.25&0.290\\
  &$K^- K^{*+}$ & $T\,'_P+E\,'_V$
     & $0.38\pm0.08$&0.43&0.43&0.431\\
  &$K^0 \ol{K}^{*0}$ & $E\,'_V-E\,'_P$
    & $<0.17$&0.08&0.16&0.052 \\
  &$\ol{K}^0 K^{*0}$ & $E\,'_P-E\,'_V$
     & $<0.09$  &0.08&0.16&0.062 \\
  &$\pi^0 \phi$ & $\frac{1}{\sqrt{2}}(C\,'_P+SE\,'_P)$
     & $<0.14$&0.12& 0.12&0.105 \\
  &$\ol{K}^{*0} \eta\,'$ & {\small $-\frac{1}{\sqrt{6}}(C_P + E_P + 2
  E_V+4SE_V)$}
  & $< 0.10$&0.004&0.003 &  \\
  $D^0$&$\eta \phi$ &
     $\frac{1}{\sqrt{3}}(C\,'_P-2SE\,'_P+SE\,'_V)$
     & $<2.8$&0.035&0.034& \\
 &$\pi^+ \rho^-$ & $-(T\,'_V+E\,'_P)$
     & --- &0.34& 0.35&0.485 \\
  &$\pi^- \rho^+$ &  $-(T\,'_P+E\,'_V)$
     & --- &0.62&0.61 & 0.706\\
  &$\pi^0 \rho^0$ &  $\frac12(C\,'_P+C\,'_V-E_P-E_V)$
     & --- &0.19& 0.16& 0.216 \\
  &$\pi^0 \omega$ &{\small
  $\frac12(C\,'_V-C\,'_P+E\,'_P+E\,'_V+2SE\,'_P)$}
     & --- &0.020&0.003 & 0.013\\
  &$\eta \omega$ &{\small
$-\frac{1}{\sqrt{6}}(C\,'_P+2C\,'_V+SE\,'_V+4SE\,'_P)$}
     & --- &0.13&0.10&  \\
  &$\eta\,' \omega$ &{\small
  $\frac{1}{2\sqrt{3}}(C\,'_P-C\,'_V+4SE\,'_V-2SE\,'_P)$}
     & ---&0.0007&0.0003&  \\
  &$\eta \rho^0$ &{\small
  $\frac{1}{\sqrt{6}}(2C\,'_V-C\,'_P-SE\,'_V)$}
     & ---&0.0039 &0.0015& \\
  &$\eta\,' \rho^0$ & $\frac{1}{2\sqrt{3}}(C\,'_V+C\,'_P+4SE\,'_V)$
     & --- &0.012& 0.009& 0.039\\
  &$K^{*+}\pi^-$& $-(T\pp_P+E\pp_V)$&---&0.029&0.029&0.025\\
  &$K^+\rho^-$&$ -(T\pp_V+E\pp_P)$&---&0.016&0.016& 0.004\\
  &$K^{*0}\pi^0$&$ -\frac{1}{\sqrt{2}}(C\pp_P-E\pp_V)$&---&0.0052&0.0064&0.008 \\
  &$K^0\rho^0$&$-\frac{1}{\sqrt{2}}(C\pp_V-E\pp_P)$&---&0.0069&0.0059&  \\
  &$K^{*0}\eta$&{\small $ -\frac{1}{\sqrt{3}}(C\pp_P-E\pp_P+E\pp_V+SE\pp_V)$}&
  ---&0.0030&0.0041&  \\
  &$K^{*0}\eta'$&{\small $\frac{1}{\sqrt{6}}(C\pp_P+2E\pp_P+E\pp_V+4SE\pp_V)$}&
  ---&0.0&0.0& \\
  &$K^0\omega$&$\frac{1}{\sqrt{2}}(C\pp_V+E\pp_P)$&---&0.0076&0.0056&0.002 \\
  &$K^0\phi$&$E\pp_V+SE\pp_P$&---&0.0&0.0006& \\
\hline
\end{tabular}
\end{center}
\end{table}
\addtocounter{table}{-1}
\begin{table}[hbp]
\caption{(\em continued).} \vspace{8pt}
\begin{center}
\begin{tabular}{|c|c|c|c|c|c|c|}
\hline  \multicolumn{2}{|c|}{Decay Modes}& Representation
     & Experimental &\multicolumn{2}{|c|}{Present ${\cal B} (\times 10^{-2})$}& LP \cite{lp2002}\\
     \cline{5-6}
 \multicolumn{2}{|c|}{}& &${\cal B} (\times 10^{-2})$&FIT A&FIT B&${\cal B} (\times 10^{-2})$\\
\hline
 & $\ol{K}^{*0} \pi^+$ & $T_V+C_P$
     & $1.92 \pm 0.19$ &1.96&1.96&1.996\\
 & $\pi^+ \phi$  &$C\,'_P-SA\,'_P$& $0.61\pm 0.06$
  &0.64&0.62 &0.619\\
 &  $\ol{K}^0 \rho^+$ & $T_P+C_V$
     & $6.6 \pm 2.5$&7.56 &8.43& 12.198\\
 &  $\pi^+ \rho^0$ &{\small
   $-\frac{1}{\sqrt{2}}(T\,'_V+C\,'_P-A\,'_P+A\,'_V)$}
     & $0.104\pm0.018$ &0.088&0.088&0.104 \\
  &$K^+ \ol{K}^{*0}$ & $T\,'_V-A\,'_V$
     & $0.42\pm0.05$ &0.44&0.44&0.436 \\
  &$\ol{K}^0 K^{*+}$ &$T\,'_P-A\,'_P$
     & $3.1\pm1.4$&1.43 &1.25&  1.515\\
 &$K^+\rho^0$&$-\frac{1}{\sqrt{2}}(C\pp_V-A\pp_P)$&$0.025 \pm 0.012$&0.030&0.025&
 0.029\\
  &$K^{*0}\pi^+$&$ -(C\pp_P+A\pp_V)$&$0.036 \pm 0.016$ &0.024&0.022 &0.027 \\
 $D^+$ &$K^+\phi$&$ -(A\pp_V+SA\pp_P)$&$<0.013$&0.0066&0.0067 & \\
  &$\pi^+\omega$&{\small $\frac{1}{\sqrt{2}}(T\,'_V+C\,'_P+A\,'_V+A\,'_P+2SA\,'_P)$}&
  ---&0.57&0.58& \\
  &$\eta\rho^+$&{\small $\frac{1}{\sqrt{3}}(T\,'_P+2C\,'_V+A\,'_V+A\,'_P+SA\,'_V)$}&---&0.24&0.43& \\
  &$\eta'\rho^+$&{\small $-\frac{1}{\sqrt{6}}(T\,'_P-C\,'_V+A\,'_V+A\,'_P+4SA\,'_V)$}&---&0.15&0.15&
  \\
  &$\pi^0 \rho^+$ &{\small
$-\frac{1}{\sqrt{2}}(T\,'_P+C\,'_V+A\,'_P-A\,'_V)$}
     & ---&0.28&0.35& 0.451 \\
  &$K^0\rho^+$&$-(C\pp_V+A\pp_P)$&---&0.025&0.022& 0.042 \\
  &$\pi^0K^{*+}$&$-\frac{1}{\sqrt{2}}(C\pp_P-A\pp_V)$&---&0.037&0.036&0.057 \\
  &$K^+\omega$&$-\frac{1}{\sqrt{2}}(C\pp_V+A\pp_P)$&---&0.012&0.011& \\
  &$K^{*+}\eta$&{\small
  $\frac{1}{\sqrt{3}}(T\pp_P-A\pp_P+A\pp_V+SA\pp_V)$}&---&0.015&0.015
  & \\
  &$K^{*+}\eta'$&{\small
  $-\frac{1}{\sqrt{6}}(T\pp_P+2A\pp_P+A\pp_V+4SA\pp_V)$}&---
  &0.00014&0.00016 & \\
\hline &  $\ol{K}^{*0} K^+$ & $C_P+A_V$
     & $3.3 \pm 0.9$&3.34 &3.42&4.812\\
 & $\ol{K}^0 K^{*+}$ & $C_V+A_P$
     & $4.3 \pm 1.4$&4.98 &4.66&2.467 \\
  &$\pi^+ \rho^0$ & $\frac{1}{\sqrt{2}}(A_V-A_P)$
     &$0.06^{\ddagger}(<0.07)$&0.06 &0.06 &\\
 & $\pi^+ \phi$ & $T_V+SA_P$
     & $3.6 \pm 0.9$ &3.08& 2.93& 4.552\\
  &$\pi^+ K^{*0}$ &  $-(T\,'_V-A\,'_V)$
     & $0.65\pm0.28$&0.33&0.35 & 0.445 \\
  &$K^+ \rho^0$ &$-\frac{1}{\sqrt{2}}(C\,'_P+A\,'_P)$
     & $<0.29$ &0.12& 0.12& 0.198 \\
  $D^+_s$ &   $K^+\phi$&$T\,'_V+C\,'_P+A\,'_V+SA\,'_P$&$<0.05$ &0.032& 0.033&0.008 \\
 & $K^+ \omega$ &$-\frac{1}{\sqrt{2}}(C\,'_P-A\,'_P-2SA\,'_P)$
     & ---&0.40&0.39 & 0.178\\
  &$K^0 \rho^+$ &$-(T\,'_P-A\,'_P)$
     & --- &0.91&0.77& 1.288\\
  &$\pi^0 K^{*+}$ & $-\frac{1}{\sqrt{2}}(C\,'_V+A\,'_V)$
     & ---&0.13 & 0.13&  0.076 \\
  &$\eta K^{*+}$&{\small $\frac{1}{\sqrt{3}}(T\,'_P+2C\,'_V+A\,'_P-A\,'_V-SA\,'_V)$}&---&0.038&0.047&0.146\\
  &$\eta\,' K^{*+}$&{\small
  $\frac{1}{\sqrt{3}}(2T\,'_P+C\,'_V+2A\,'_P+A\,'_V+4SA\,'_V)$}&---
  &0.068&0.059&\\
  &$K^{*0}K^+$&$-(T\pp_V+C\pp_P)$&---&0.0015&0.0015& 0.006\\
  &$K^{*+}K^0$&$-(T\pp_P+C\pp_P)$&---&0.0076&0.0085& 0.018\\
\hline
\end{tabular}
\end{center}
\begin{center}
\parbox{12cm}{
\small \baselineskip=1.0pt \hspace{-2cm}$^{\ddagger}$   The
central value of the E791 experiment \cite{e791}.}
\end{center}
\end{table}
\begin{table}[hbp]
\caption{\small SU(3) flavor symmetry relations of $D\to PP$ decay
modes and breaking of the relations.
$\lambda=|V^*_{cs}V_{us}/V^*_{cs}V_{ud}|\approx 0.226$.
$\kappa=|V^*_{cs}V_{us}/V^*_{cd}V_{us}|\approx 4.446$.
\label{tab:breaking1}} \vspace{8pt} \tabcolsep0.5in
\begin{tabular}{|c|c|c|}
\hline SU(3) Symmetry Relations&\multicolumn{2}{|c|}{LHS of Relations}\\
\cline{2-3}
                               & FIT $\alpha$&FIT $\beta$\\
\hline \hline $\frac{|{\cal A}(D^0\to
\pi^+\pi^-)+\sqrt{2}{\cal A}(D^0\to \pi^0\pi^0)|}{|\sqrt{2}{\cal A}(D^+\to \pi^+\pi^0)|}=1$&1.00&1.00\\
\hline \hline  $\frac{|{\cal A}(D^0\to
K^-\pi^+)+\sqrt{2}{\cal A}(D^0\to \ol{K}^0\pi^0)|}{|{\cal A}(D^+\to \ol{K}^0\pi^+)|}=1$&1.00&1.00\\
\hline \hline  $\frac{|\lambda{\cal A}(D^+\to
\pi^+\ol{K}^0)+\kappa{\cal
A}(D^+\to K^0\pi^+)|}{|\sqrt{2}\kappa{\cal A}(D^+\to K^+\pi^0)|}=1$&0.49&0.79\\
\hline  $\frac{|\lambda{\cal A}(D^+\to
\ol{K}^0\pi^+)+\sqrt{2}\kappa{\cal
A}(D^+\to K^+\pi^0)|}{|\kappa{\cal A}(D^+\to K^0\pi^+)|}=1$&1.56&1.11\\
\hline  $\frac{|\sqrt{2}\kappa{\cal A}(D^+\to \pi^0
K^+)-\kappa{\cal
A}(D^+\to K^0\pi^+)|}{|\lambda{\cal A}(D^+\to \ol{K}^0\pi^+)|}=1$&2.21&1.82\\
\hline \hline $\frac{|{\cal A}(D^0\to K^+K^-)|}{|\kappa{\cal
A}(D^0\to K^+\pi^-)|}=1$&1.27&1.24\\
\hline  $\frac{|\kappa{\cal A}(D^0\to K^+\pi^-)|}{|{\cal
A}(D^0\to \pi^+\pi^-)|}=1$&1.43&1.43\\
\hline  $\frac{|\kappa{\cal A}(D^0\to K^+\pi^-)|}{|\lambda{\cal
A}(D^0\to K^-\pi^+)|}=1$&1.20&1.24\\
\hline \hline $\frac{|\lambda{\cal A}(D^0\to
\ol{K}^0\pi^0)|}{|{\cal
A}(D^0\to \pi^0\pi^0)|}=1$&1.26&1.12\\
\hline  $\frac{|\lambda{\cal A}(D^0\to
\ol{K}^0\pi^0)|}{|\kappa{\cal
A}(D^0\to K^0\pi^0)|}=1$&1.78&1.10\\
\hline \hline $\frac{|{\cal A}(D^+\to
K^+\ol{K}^0)|}{|\sqrt{2}\kappa{\cal
A}(D^+\to K^+\pi^0)|}=1$&0.89&0.86\\
\hline  $\frac{|\sqrt{2}\kappa{\cal A}(D^+\to K^+\pi^0)|}{|{\cal
A}(D^+_s\to K^0\pi^+)|}=1$&1.24&1.24\\
\hline \hline $\frac{|\lambda{\cal A}(D^+_s\to
\ol{K}^0K^+)|}{|\sqrt{2}{\cal
A}(D^+_s\to K^+\pi^0)|}=1$&1.34&0.98\\
\hline $\frac{|\kappa{\cal A}(D^+\to K^0\pi^+)|}{|\lambda{\cal
A}(D^+_s\to K^+\pi^0)|}=1$&1.08&1.14\\
\hline \hline $\frac{|\lambda{\cal A}(D^+\to
\ol{K}^0\pi^+)|}{|\sqrt{2}{\cal
A}(D^+\to \pi^0\pi^+)|}=1$&0.55&0.67\\
\hline
\end{tabular}
\end{table}
\begin{table}[hbp]
\caption{\small SU(3) flavor symmetry relations of $D \to PV$
decays and breaking of the relations.
$\lambda=|V^*_{cs}V_{us}/V^*_{cs}V_{ud}|\approx 0.226$.
$\kappa=|V^*_{cs}V_{us}/V^*_{cd}V_{us}|\approx 4.446$.
\label{tab:breaking2}} \vspace{8pt} \tabcolsep0.5in
\begin{tabular}{|c|c|c|}
\hline SU(3) Symmetry Relations&\multicolumn{2}{|c|}{LHS of Relations}\\
\cline{2-3}                    &FIT A&FIT B \\
\hline \hline $\frac{|{\cal A}(D^0\to \pi^+K^{*-})+\sqrt{2}{\cal
A}(D^0\to \pi^0\ol {K}^{*0})|}{|{\cal A}(D^+\to \pi^+\ol
{K}^{*0})|}=1$&1.00&1.00\\
\hline \hline $\frac{|{\cal A}(D^0\to \rho^+K^-)+\sqrt{2}{\cal
A}(D^0\to \rho^0\ol {K}^0)|}{|{\cal A}(D^+\to \ol
{K}^0\rho^+)|}=1$&1.00&1.00\\
\hline \hline $\frac{|{\cal A}(D^0\to \ol
{K}^0\phi)-{\cal A}(D^+_s\to \pi^+\phi)|}{|{\cal A}(D^0\to \pi^+K^{*-})|}=1$&1.00&1.00 \\
\hline $\frac{|{\cal A}(D^0\to \pi^+K^{*-})+{\cal A}(D^0\to \ol
{K}^0\phi)|}{|{\cal A}(D^+_s\to \pi^+\phi)|}=1$&0.99& 0.99 \\
\hline $\frac{|{\cal A}(D^0\to \pi^+K^{*-})-{\cal A}(D^+_s\to
\pi^+\phi)|}{|{\cal A}(D^0\to \ol{K}^0\phi)|}=1$&1.00&1.00\\
\hline \hline $\frac{|\lambda\sqrt{2}{\cal A}(D^+_s\to
\pi^+\rho^0)+\sqrt{2}{\cal A}(D^+\to \pi^+\rho^0)|}{|\lambda{\cal
A}(D^+\to \pi^+\ol{K}^{*0})|}=1$&0.88&0.88 \\
\hline $\frac{|\lambda{\cal A}(D^+\to
\pi^+\ol{K}^{*0})+\sqrt{2}{\cal A}(D^+\to
\pi^+\rho^0)|}{|\lambda\sqrt{2}{\cal A}(D^+_s\to
\pi^+\rho^0)|}=1$&0.60&0.59 \\
\hline $\frac{|\lambda{\cal A}(D^+\to
\pi^+\ol{K}^{*0})+\lambda\sqrt{2}{\cal A}(D^+_s\to
\pi^+\rho^0)|}{|\sqrt{2}{\cal A}(D^+\to
\pi^+\rho^0)|}=1$&1.10&1.10\\
\hline \hline $\frac{|\lambda\sqrt{2}{\cal A}(D^+_s\to
\pi^+\rho^0)-\sqrt{2}{\cal A}(D^+\to \pi^0\rho^+)|}{|\lambda{\cal
A}(D^+\to \rho^+\ol{K}^{0})|}=1$&1.03&1.03\\
\hline $\frac{|\lambda{\cal A}(D^+\to
\rho^+\ol{K}^{0})+\sqrt{2}{\cal A}(D^+\to
\pi^0\rho^+)|}{|\lambda\sqrt{2}{\cal A}(D^+_s\to
\pi^+\rho^0)|}=1$&1.48&1.10 \\
\hline $\frac{|\lambda\sqrt{2}{\cal A}(D^+_s\to
\pi^+\rho^0)|-\lambda{\cal
A}(D^+\to \rho^+\ol{K}^{0})}{|\sqrt{2}{\cal A}(D^+\to \pi^0\rho^+)|}=1$&0.95&0.97\\
\hline \hline $\frac{|\lambda{\cal A}(D^+_s\to K^+\ol
{K}^{*0})+{\cal A}(D^+\to K^+\ol {K}^{*0})|}{|\lambda{\cal
A}(D^+\to \pi^+\ol
{K}^{*0})|}=1$&0.58&0.57 \\
\hline $\frac{|\lambda{\cal A}(D^+\to \pi^+\ol {K}^{*0})-{\cal
A}(D^+\to K^+\ol {K}^{*0})|}{|\lambda{\cal
A}(D^+_s\to K^+\ol {K}^{*0})|}=1$&1.17&1.16 \\
\hline $\frac{|\lambda{\cal A}(D^+_s\to K^+\ol
{K}^{*0})-\lambda{\cal A}(D^+\to \pi^+\ol {K}^{*0})|}{|{\cal
A}(D^+\to K^+\ol {K}^{*0})|}=1$&0.96&0.97\\
\hline \hline $\frac{|\lambda{\cal A}(D^+_s\to \ol
{K}^0K^{*+})+{\cal A}(D^+\to \ol {K}^0K^{*+})|}{|\lambda{\cal
A}(D^+\to \rho^+\ol
{K}^{0})|}=1$&1.09&1.19 \\
\hline $\frac{|\lambda{\cal A}(D^+\to \rho^+\ol {K}^{0})-{\cal
A}(D^+\to \ol {K}^0K^{*+})|}{|\lambda{\cal
A}(D^+_s\to \ol {K}^0K^{*+})|}=1$&1.01&0.97\\
\hline $\frac{|\lambda{\cal A}(D^+_s\to \ol
{K}^0K^{*+})-\lambda{\cal A}(D^+\to \rho^+\ol {K}^{0})|}{|{\cal
A}(D^+\to \ol {K}^0K^{*+})|}=1$&0.98&0.96\\
\hline \hline $\frac{|\lambda{\cal A}(D^+_s\to \ol
{K}^0K^{*+})+\sqrt{2}{\cal A}(D^+_s\to
K^{*+}\pi^0)|}{|\lambda\sqrt{2}{\cal A}(D^+_s\to
\pi^+\rho^0)|}=1$&1.42&1.31\\
\hline $\frac{|\lambda\sqrt{2}{\cal A}(D^+_s\to
\pi^+\rho^0)+\sqrt{2}{\cal A}(D^+_s\to
K^{*+}\pi^0)|}{|\lambda{\cal A}(D^+_s\to \ol
{K}^0K^{*+})|}=1$&0.92&0.95\\
\hline $\frac{|\lambda\sqrt{2}{\cal A}(D^+_s\to
\pi^+\rho^0)+\lambda{\cal A}(D^+_s\to \ol
{K}^0K^{*+})|}{|\sqrt{2}{\cal A}(D^+_s\to
K^{*+}\pi^0)|}=1$&1.11&1.07\\
\hline \hline
$\frac{|{\cal A}(D^+_s\to K^0\rho^+)|}{|{\cal A}(D^+\to \ol{K}^0K^{*+})|}=1$&0.91&0.89 \\
\hline \hline
$\frac{|{\cal A}(D^+_s\to \pi^+K^{*0})|}{|{\cal A}(D^{+}\to K^{+}\ol {K}^{*0})|}=1$&0.93&0.94 \\
\hline \hline
$\frac{|{\cal A}(D^0\to K^+K^{*-})|}{|\kappa {\cal A}(D^0\to K^+\rho^-)|}=1$&1.05&1.05\\
\hline $\frac{|\lambda{\cal A}(D^0\to \pi^+K^{*-})|}{|{\cal
A}(D^0\to \pi^+\rho^-)|}=1$&1.05&1.05\\
\hline
$\frac{|{\cal A}(D^0\to K^+K^{*-})|}{|\lambda{\cal A}(D^0\to \pi^+K^{*-})|}=1$&1.14&1.14\\
\hline \hline
$\frac{|{\cal A}(D^0\to K^-K^{*+})|}{|\kappa {\cal A}(D^0\to \pi^-K^{*+})|}=1$&1.08&1.08\\
\hline $\frac{|\lambda{\cal A}(D^0\to K^-\rho^+)|}{|{\cal
A}(D^0\to \pi^-\rho^+)|}=1$&1.09&1.09\\
\hline
$\frac{|{\cal A}(D^0\to K^-K^{*+})|}{|\lambda{\cal A}(D^0\to K^-\rho^+)|}=1$&1.08&1.08\\
\hline
\end{tabular}
\end{table}


\begin{thebibliography}{999}
\bibitem{cc} L.L. Chau and H.Y. Cheng, Phys. Rev. Lett.
{\bf 56}, 1655 (1986); Phys. Rev. D{\bf 36}, 137 (1987); Phys.
Lett. B{\bf 222}, 285 (1989); Mod. Phys. Lett. A{\bf 4}, 877
(1989); L.L. Chau, Phys. Rep. {\bf 95}, 1 (1983).
\bibitem{bsw} M. Bauer, B. Stech and M. Wirbel, Z. Phys. C{\bf
34}, 103 (1987).
\bibitem{history} V. Barger and S. Pakvasa, Phys. Rev. Lett. {\bf
43}, 812 (1979); H.J. Lipkin, Phys. Rev. Lett. {\bf 44}, 710
(1980); X.Y. Pham, Phys. Lett. B{\bf 193}, 331 (1987); J.G.
Korner, K. Schilcher, M. Wirbel and Y.L. Wu, Z. Phys. C {\bf 48},
663 (1990); F. Buccella, M. Lusignoli, G. Miele, A. Pugliese and
P. Santorelli, Phys. Rev. D{\bf 51}, 3478 (1995); F. Buccella, M.
Lusignoli and A. Pugliese, Phys. Lett. B{\bf 379}, 249 (1996); M.
Ablikim, D.S. Du and M.Z. Yang, Phys. Lett. B{\bf 536}, 34 (2002).
\bibitem{bfbb} S. Bianco, F.L. Fabbri, D. Benson and I. Bigi, hep-ex/0309021.
\bibitem{cpv} S.E. Csorna etal, CLEO Collaboration, Phys. Rev. D{\bf 65}, 092001 (2002);
B.D. Yabsley, hep-ex/0311057; A.A. Petrov, hep-ph/0403030.
\bibitem{fsi} X.Q. Li and B.S. Zou, Phys. Lett. B{\bf 399}, 297
(1997); Y.S. Dai, D.S. Du, X.Q. Li, Z.T. Wei and B.S. Zou, Phys.
Rev. D{\bf 60}, 014014 (1999); S. Fajfer and J. Zupan, Int. J.
Mod. Phys. A{\bf 14}, 4161 (1999); P. Zenczykowski, Phys. Lett.
B{\bf 460}, 390 (1999); J.O. Eeg, S. Fajfer and J. Zupan, Phys.
Rev. D{\bf 64}, 034010 (2001); J.W. Li, M.Z. Yang and D.S. Du,
hep-ph/0206154; M. Ablikim, D.S. Du and M.Z. Yang, hep-ph/0211413;
S. Fajfer, A. Prapotnik, P. Singer and J. Zupan, Phys. Rev. D{\bf
68}, 094012 (2003).
\bibitem{bbns} M. Beneke, G. Buchalla, M. Neubert and C.T.
Sachrajda, Phys. Rev. Lett. {\bf 83}, 1914 (1999).
\bibitem{lixn} H.N. Li and H.L. Yu, Phys. Rev. Lett. {\bf 74},
4388 (1995).
\bibitem{kv} A.N. Kamal and R.C. Verma, Phys. ReV. D{\bf 35}, 3515
(1987); ERRATUM-ibid. D{\bf 36}, 3527 (1987); R.C. Verma and A.N.
Kamal, Phys. Rev. D{\bf 43}, 829 (1991).
\bibitem{rosner} J.L. Rosner, Phys. Rev. D{\bf 60}, 114026 (1999).
\bibitem{chiang} C.W. Chiang, Z. Luo, and J.L.
Rosner, Phys. Rev. D{\bf 67}, 014001 (2003).
\bibitem{ghlr} M. Gronau, O.F. Hern\'{a}ndez, D. London and J.L.
Rosner, Phys. Rev. D{\bf 50}, 4529 (1994); ibid. {\bf 52}, 6356,
6374 (1995).
\bibitem{bre1} L.L. Chau and H.Y. Cheng, Phys. Lett. B{\bf 333},
514 (1994).
\bibitem{bre2} M. Gronau and D. Pirjol, Phys. Rev.
D{\bf 62}, 077301 (2000).
\bibitem{zww2003} M. Zhong, Y.L. Wu and W.Y. Wang, Eur. Phys. J. C
(2003), \\DOI 10.1140/epjcd/s2003-03-017-1.
\bibitem{cl2002} F.E. Close and H.J. Lipkin, Phys. Lett. B{\bf
551}, 337 (2003).
\bibitem{gronau} M. Gronau and J.L. Rosner, Phys. Rev. D{\bf 53},
2516 (1996); A.S. Dighe, M. Gronau and J.L. Rosner, Phys. Lett.
B{\bf 367}, 357 (1996); ibid. {\bf 377}, 325(E) (1996).
\bibitem{feldmann} T. Feldmann and P. Kroll, Eur. Phys. J. C{\bf
5}, 327 (1998); T. Feldmann, P. Kroll and B. Stech, Phys. Rev.
D{\bf 58}, 114006 (1998); Phys. Lett. B{\bf 449}, 339 (1999); T.
Feldmann and P. Kroll, Phys. Scripta T{\bf 99}, 13 (2002).
\bibitem{chau} L.L. Chau and H.Y. Cheng, Phys. Rev. D{\bf 39},
2788 (1989); L.L. Chau, H.Y. Cheng and T. Huang, Zeit. Phys. C{\bf
53}, 413 (1992); H.Y. Cheng and B. Tseng, Phys. Rev. D{\bf 59},
014034 (1999).
\bibitem{gfa} M. Neubert, V. Riekert, Q.P. Xu and B. Stech, in {\it Heavy Flavors},
edited by A.J. Buras and H. Lindner (World Scientific, Singapore,
1992); H.Y. Cheng, Phys. Lett. B{\bf 335}, 428 (1994).
\bibitem{cheng2003} H.Y. Cheng, Eur. Phys. J. C{\bf 26}, 551 (2003).
\bibitem{zenc} P. Zenczykowski, Acta Phys. Polon. B{\bf 28},
1605 (1997).
\bibitem{kr} A. Khodjamirian and R. R\"{u}ckl, Adv. Ser. Direct
High Energy Phys. {\bf 15}, 345 (1998).
\bibitem{ww} W.Y. Wang and Y.L. Wu, Phys. Lett. B{\bf 515}, 57 (2001);
ibid. {\bf 519}, 219 (2001); \\
M. Zhong, Y.L. Wu and W.Y. Wang, Int. J. Mod. Phys A{\bf 18}, 1959
(2003);\\
W.Y. Wang, Y.L. Wu and M. Zhong, J. Phys. G{\bf 29}, 2743 (2003).
\bibitem{flynn} J.M. Flynn and C.T. Sachrajda, Adv. Ser. Direct.
High Energy Phys. {\bf 15}, 402 (1998).
\bibitem{abada} A. Abada, D. Becirevic, Ph. Boucaud, J.P. Leroy,
V. Lubicz and F. Mescia, Nucl. Phys. B {\bf 619}, 565 (2001).
\bibitem{isgur} D. Scora and N. Isgur, Phys. Rev. D{\bf 52},
2783 (1995).
\bibitem{melikhov} D. Melikhov, Phys. Rev. D{\bf 53}, 2460 (1996);
ibid. {\bf 56}, 7089 (1997).
\bibitem{bbs} M. Wirbel, B. Stech and M. Bauer, Z. Phys. C{\bf
29}, 637 (1985).
\bibitem{ms2000} D. Melikhov and B. Stech, Phys. Rev. D{\bf 62},
014006 (2000).
\bibitem{gus} S. Gusken et al, Nucl. Phys. (Proc. Suppl.){\bf 47},
485 (1996).
\bibitem{fs1998} J. M. Flynn and C. T. Sachrajda, Adv. Ser.
Direct. High Energy Phys. {\bf 15}, 402 (1998).
\bibitem{ball} P. Ball, V. Braun and H. Dosch, Phys. Lett. B{\bf
273}, 316 (1991); Phys. Rev. D{\bf 44}, 3567 (1991); P. Ball, Phys.
Rev. D{\bf 48}, 3190 (1993).
\bibitem{wwz} W.Y. Wang, Y.L. Wu and M. Zhong, Phys. Rev. D{\bf
67}, 014024 (2003).
\bibitem{akl} A. Ali, G. Kramer and Cai-Dian L\"{u}, Phys. Rev.
D{\bf 58}, 094009 (1998).
\bibitem{ns} M. Neubert and B. Stech, Adv. Ser. Direct. High
Energy Phys. {\bf 15}, 294 (1998).
\bibitem{PDG2002}
K.~Hagiwara {\it et al.}, Phys. Rev. D{\bf 66}, 010001 (2002)
(URL: http://pdg.lbl.gov).
\bibitem{bbl} G. Buchalla, A.J. Buras and M.E. Lautenbacher, Rev.
Mod. Phys.{\bf 68}, 1125 (1996).
\bibitem{e791} Fermilab E791 Collabortion, E. M. Aitala et al.,
Phys. Rev. Lett.{\bf 86}, 765 (2001).
\bibitem{lp2002} M. Lusignoli and A. Pugliese, hep-ph/0210071.
\end{thebibliography}
\end {document}